\documentclass[fleqn,twoside]{article}
\usepackage{espcrc2}

\usepackage{epsfig}
\usepackage[dvips]{color}
\usepackage{latexsym}
\title{Topological susceptibility with the improved Asqtad action
}

\author{ MILC Collaboration: C.~Bernard
\address{Department of Physics, Washington University, St.~Louis, MO 63130, USA},
T.~Burch
\address{Department of Physics, University of Arizona, Tucson, AZ 85721, USA}, 
T.A.~DeGrand
\address{Physics Department, University of Colorado, Boulder, CO 80309, USA},
C.E.~DeTar
\address{Physics Department, University of Utah, Salt Lake City, UT
  84112, USA}\thanks{Presented by C.~DeTar.},
Steven~Gottlieb
\address{Department of Physics, Indiana University, Bloomington, IN 47405, USA and Fermilab, Batavia, IL 60510, USA},
E.~Gregory$\,\null^{\rm b}$,
A.~Hasenfratz$\,\null^{\rm c}$,
U.M.~Heller
\address{CSIT, Florida State University, Tallahassee, FL 32306-4120, USA},
J.~Hetrick
\address{University of the Pacific, Stockton, CA 95211, USA},
J.~Osborn$\,\null^{\rm d}$,
R.L.~Sugar
\address{Department of Physics, University of California, Santa Barbara, CA 93106, USA},
and D.~Toussaint$\,\null^{\rm b}$
} %end \author
\begin{document}

\begin{abstract}
As a test of the chiral properties of the improved Asqtad (staggered
fermion) action, we have been measuring the topological susceptibility
as a function of quark masses for 2 + 1 dynamical flavors.  We report
preliminary results, which show reasonable agreement with leading
order chiral perturbation theory for lattice spacing less than 0.1 fm.
The total topological charge, however, shows strong persistence over
Monte Carlo time.
\end{abstract}

\maketitle

\section{INTRODUCTION}

The cost of lattice simulations increases dramatically as the lattice
spacing is decreased.  Improvement programs seek to avoid these costs
by reducing lattice artifacts at larger lattice spacing.  Whether
lattice artifacts are successfully reduced depends on the observable.
The staggered fermion Asqtad action \cite{ref:asqtad} is designed to
remove ${\cal O}(a^2)$ lattice artifacts at tree level, resulting in
an improved flavor symmetry and better chiral properties
\cite{ref:asqtad,ref:asqtadflavor}.  However, the Asqtad quark-gluon
vertex is not as smooth as that of the more elaborate HYP action
\cite{ref:HYP} and zero modes are not treated as rigorously as with
the more expensive domain wall and overlap actions.  Through a rougher
vertex, quark propagation might be influenced by small instanton-like
dislocations, and with imprecise zero modes, at small quark mass the
fermion determinant may fail to suppress adequately configurations
with nonzero topological charge.

\section{TOPOLOGICAL SUSCEPTIBILITY}
A good test of the effectiveness of zero mode suppression is the
identity in chiral perturbation theory \cite{ref:topochiral} relating
the susceptibility of the topological charge to the light quark masses
\begin{eqnarray}
 \chi_{\rm top} = \frac{\langle Q^2 \rangle}{V} = 
  \frac{\Sigma}{\left( \frac{1}{m_u} + \frac{1}{m_d} + \frac{1}{m_s}\right)},
\end{eqnarray}
where $Q =  \int F^a_{\mu\nu} \tilde F^a_{\mu\nu} d^4x/32 \pi^2.$
For three flavors with $m_u = m_d$ we write the susceptibility as
\begin{equation}
 \chi_{\rm top} = f_\pi^2 m_\pi^2/[4(1 + m_{ud}/2 m_s)],
\label{eq:chiral}
\end{equation}
\begin{figure}[h]
 \hspace*{-12mm}
 \epsfig{bbllx=100,bblly=230,bburx=730,bbury=740,clip=,
         file=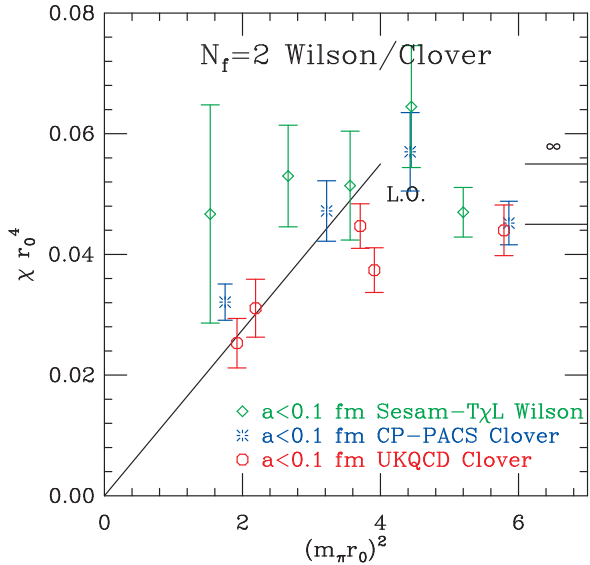,width=125mm}
 \vspace*{-35mm}
 \caption{Topological susceptibility for the unimproved and improved
 Wilson action {\it vs.} pion mass squared in units of $r_0$
 \protect\cite{ref:previousWilson}.  Lines at the right indicate the
 quenched value.}
 \label{fig:wilson}
 \vspace*{-7mm}
\end{figure}
showing that it vanishes linearly in the square of the pion mass in
the chiral limit.

\begin{figure}[h]
 \hspace*{-10mm}
 \epsfig{bbllx=100,bblly=230,bburx=730,bbury=740,clip=,
         file=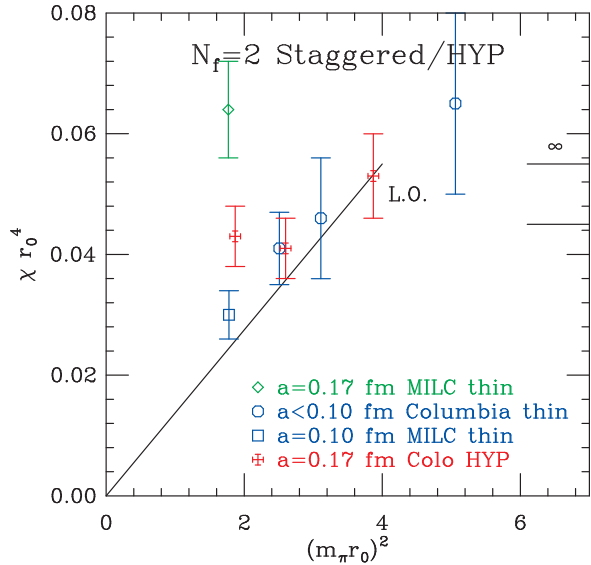,width=125mm}
 \vspace*{-35mm}
 \caption{Topological susceptibility for the unimproved thin-link and
 improved HYP staggered fermion action {\it vs.} pion mass squared in
 units of $r_0$ \protect\cite{ref:HYPpreserve,ref:dynHYP02}. Lines at
 the right indicate the quenched value.}
 \vspace*{-7mm}
 \label{fig:staggered}
\end{figure}

Shown in Figs.~\ref{fig:wilson} and \ref{fig:staggered} are recent
results for the $N_f = 2$ relation, for the unimproved and improved
Wilson actions and the conventional thin-link staggered fermion action
\cite{ref:HYP,ref:previousWilson,ref:HYPpreserve}, suggesting that at
a lattice spacing $a < 0.1$ fm the chiral prediction \ref{eq:chiral}
is reasonably well satisfied as far as testing was possible.  More
recently, as shown in Fig.~\ref{fig:staggered} the relation was tested
in simulations with dynamical HYP fermions at $a = 0.17$ fm with a
dramatic improvement over thin-link, but suggesting that even here a
smaller lattice spacing is needed \cite{ref:dynHYP02}.

\section{MEASUREMENT}

We have measured the topological charge on an ensemble of $20^3\times 64$
($a = 0.13$ fm) and $28^3\times 96$ ($a = 0.09$ fm) gauge configurations
generated in the presence of $2+1$ flavors of Asqtad dynamical quarks
of varying masses.  The charge is obtained by integrating the
topological charge density, defined as an approximation to
$F^a_{\mu\nu} \tilde F^a_{\mu\nu}/32 \pi^2$.  
\begin{figure}[h]
 \hspace*{-10mm}
 \epsfig{bbllx=100,bblly=230,bburx=730,bbury=740,clip=,
         file=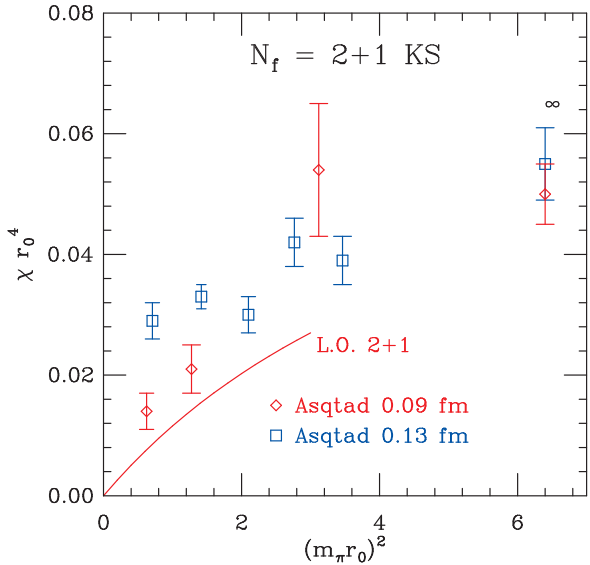,width=125mm}
 \vspace*{-35mm}
 \caption{Topological susceptibility {\it vs.} pion mass squared for the
 Asqtad action with $2+1$ flavors.  The curve gives the prediction of
 leading order chiral perturbation theory.  Quenched values are
 plotted on the extreme right.}
 \vspace*{-9mm}
 \label{fig:asqtadsuscept}
\end{figure}
As usual, prior to measuring the topological charge, it is necessary
to smooth the gauge configurations to remove short wavelength
fluctuations.  Smoothing was done with a series of hypercubic blocking
sweeps \cite{ref:HYPblock}.  This method seeks to achieve a smooth
configuration with a minimal distortion of the topological structures
\cite{ref:HYPpreserve}.  Accordingly we take as few smoothing sweeps
as necessary, and find that the susceptibility is constant within
errors after from one to six such sweeps.  The improved cooling
technique gave similar values of $Q$ \cite{ref:Hetrick}.

Results are shown in Fig.~\ref{fig:asqtadsuscept}.  Evidently, the
Asqtad action, like the conventional staggered action, requires a
lattice spacing less than approximately 0.1 fm to obtain reasonably
good agreement with leading order chiral perturbation theory.

\section{PERSISTENCE OF TOPOLOGICAL CHARGE}

\begin{figure}[t]
 \hspace*{-12mm}
 \epsfig{bbllx=100,bblly=-150,bburx=730,bbury=350,clip=,
         file=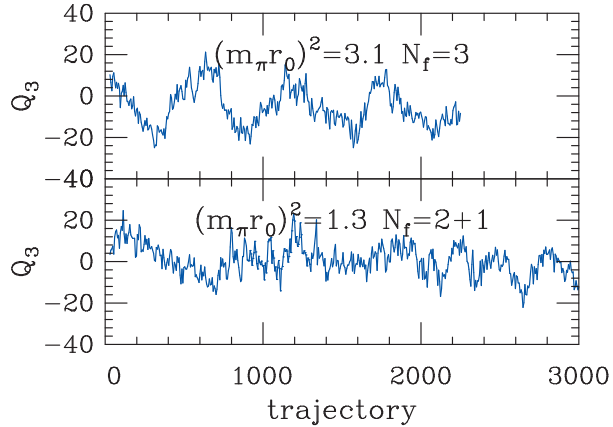,width=125mm}
 \vspace*{-50mm}
 \caption{Time history of topological charge after three HYP sweeps
 {\it vs.} trajectory at two values of $m_{u,d}$ in the $a = 0.09$ fm
 Asqtad simulation.}
 \vspace*{-9mm}
 \label{fig:persist}
\end{figure}

A particularly noteworthy feature of our measurements is the
persistence of the total topological charge.  Time histories for two
light quark masses are shown in Fig.~\ref{fig:persist}.  This
phenomenon has been noted for the conventional staggered fermion
action \cite{ref:Boyd} and appears to be rather more pronounced for
Asqtad with a better gauge action at comparable lattice spacing and
pion mass, but in our case worse for the heavier light quark.  In both
cases the R algorithm is used in the updating process.  For the pure
gauge theory, overrelaxation plus heatbath updating decorrelates
rapidly.

\section{CONCLUSIONS}

Our preliminary findings are that the Asqtad action, like the
conventional staggered fermion action, requires a lattice spacing less
than approximately 0.1 fm to give reasonably good chiral behavior for
the topological susceptibility.  However, in connection with a small
step-size updating algorithm, the action gives strong persistence of
the topological charge.  We are investigating the implications of
these results.

Computations were performed at LANL, NERSC, NCSA, ORNL, PSC, SDSC,
FNAL, the CHPC (Utah) and the Indiana University SP.  This work is
supported by the U.S. NSF and DOE.

\begin{thebibliography}{}
%%%%%%%%%%%%%%%%%%%%%%%%%%%%%%%%%%%%%%%%%%%%%%%%%%%%%%%%%%%%%%%%%%%%%%
  \bibitem{ref:asqtad}
%
K.~Orginos and D.~Toussaint,
%``Testing improved actions for dynamical Kogut-Susskind quarks,''
Phys.\ Rev.~\ D {\bf 59} (1999) 014501;
%[arXiv:hep-lat/9805009].
%%CITATION = HEP-LAT 9805009;%%
%
K.~Orginos, D.~Toussaint and R.~L.~Sugar,
%``Variants of fattening and flavor symmetry restoration,''
Phys.\ Rev.~\ D {\bf 60} (1999) 054503;
%[arXiv:hep-lat/9903032].
%%CITATION = HEP-LAT 9903032;%%
%
G.~P.~Lepage,
%``Flavor-symmetry restoration and Symanzik improvement for staggered  quarks,''
Phys.\ Rev.\ D {\bf 59} (1999) 074502.
%[arXiv:hep-lat/9809157].
%%CITATION = HEP-LAT 9809157;%%
%%%%%%%%%%%%%%%%%%%%%%%%%%%%%%%%%%%%%%%%%%%%%%%%%%%%%%%%%%%%%%%%%%%%%%
  \bibitem{ref:asqtadflavor}
%
C.~Bernard {\it et al.},
%``The QCD spectrum with three quark flavors,''
Phys.\ Rev.~\ D {\bf 64} (2001) 054506.
%[arXiv:hep-lat/0104002].
%%CITATION = HEP-LAT 0104002;%%
%%%%%%%%%%%%%%%%%%%%%%%%%%%%%%%%%%%%%%%%%%%%%%%%%%%%%%%%%%%%%%%%%%%%%%
  \bibitem{ref:HYP}
%
F.~Knechtli and A.~Hasenfratz,
%``Dynamical fermions with fat links,''
Phys.\ Rev.~\ D {\bf 63} (2001) 114502
%[arXiv:hep-lat/0012022].
%%CITATION = HEP-LAT 0012022;%%
%
and
%``Simulation of dynamical fermions with smeared links,''
hep-lat/0203010.
%%CITATION = HEP-LAT 0203010;%%
%
%%%%%%%%%%%%%%%%%%%%%%%%%%%%%%%%%%%%%%%%%%%%%%%%%%%%%%%%%%%%%%%%%%%%%%
  \bibitem{ref:topochiral}
%
H.~Leutwyler and A.~Smilga,
%``Spectrum of Dirac operator and role of winding number in QCD,''
Phys.\ Rev.~\ D {\bf 46} (1992) 5607.
%%CITATION = PHRVA,D46,5607;%%
%%%%%%%%%%%%%%%%%%%%%%%%%%%%%%%%%%%%%%%%%%%%%%%%%%%%%%%%%%%%%%%%%%%%%%
  \bibitem{ref:previousWilson}
%
G.~S.~Bali {\it et al.}  [T$\chi$L Collaboration],
%``Quark mass effects on the topological susceptibility in QCD,''
Phys.\ Rev.~\ D {\bf 64} (2001) 054502.
%[arXiv:hep-lat/0102002].
%%CITATION = HEP-LAT 0102002;%%
%
A.~Ali Khan {\it et al.}  [CP-PACS Collaboration],
%``Topological susceptibility in lattice QCD with two flavors of dynamical  quarks,''
Phys.\ Rev.~\ D {\bf 64} (2001) 114501.
%[arXiv:hep-lat/0106010].
%%CITATION = HEP-LAT 0106010;%%
%
A.~Hart and M.~Teper  [UKQCD Collaboration],
%``The topological susceptibility and f(pi) from lattice QCD,''
Phys.\ Lett.\ B {\bf 523} (2001) 280.
%[arXiv:hep-lat/0108006].
%%CITATION = HEP-LAT 0108006;%%
%
%%%%%%%%%%%%%%%%%%%%%%%%%%%%%%%%%%%%%%%%%%%%%%%%%%%%%%%%%%%%%%%%%%%%%%
  \bibitem{ref:HYPpreserve}
A.~Hasenfratz,
%``Topological susceptibility on dynamical staggered fermion configurations,''
Phys.\ Rev.~\ D {\bf 64} (2001) 074503.
%[arXiv:hep-lat/0104015].
%%CITATION = HEP-LAT 0104015;%%
%
%%%%%%%%%%%%%%%%%%%%%%%%%%%%%%%%%%%%%%%%%%%%%%%%%%%%%%%%%%%%%%%%%%%%%%
  \bibitem{ref:dynHYP02}
%
A.~Hasenfratz, this conference (2002).
%%%%%%%%%%%%%%%%%%%%%%%%%%%%%%%%%%%%%%%%%%%%%%%%%%%%%%%%%%%%%%%%%%%%%%
  \bibitem{ref:HYPblock}
%
A.~Hasenfratz and F.~Knechtli,
%``Flavor symmetry and the static potential with hypercubic blocking,''
Phys.\ Rev.~\ D {\bf 64} (2001) 034504.
%[arXiv:hep-lat/0103029].
%%CITATION = HEP-LAT 0103029;%%
%%%%%%%%%%%%%%%%%%%%%%%%%%%%%%%%%%%%%%%%%%%%%%%%%%%%%%%%%%%%%%%%%%%%%%
  \bibitem{ref:Hetrick}
J.~Hetrick, this conference (2002).
%%%%%%%%%%%%%%%%%%%%%%%%%%%%%%%%%%%%%%%%%%%%%%%%%%%%%%%%%%%%%%%%%%%%%%
  \bibitem{ref:Boyd}
%
%\cite{Boyd:1997nt}
G.~Boyd, B.~Alles, M.~D'Elia and A.~Di Giacomo,
%``Topology in QCD,''
hep-lat/9711025.
%%CITATION = HEP-LAT 9711025;%%
%%%%%%%%%%%%%%%%%%%%%%%%%%%%%%%%%%%%%%%%%%%%%%%%%%%%%%%%%%%%%%%%%%%%%%
\end{thebibliography}
\end{document}